\documentclass[aps,prl,twocolumn,preprintnumbers,showpacs]{revtex4}
\usepackage{amsmath,amssymb,bm,epsfig}
\renewcommand\d{\partial}
\newcommand\x{\mathbf{x}}

\newcommand\dlr{\raisebox{0.1em}{$\stackrel{\scriptstyle\leftrightarrow}\partial$}}
\newcommand\+{\dagger}
\renewcommand\v{\mathbf{v}}

\renewcommand\L{\mathcal{L}}

\preprint{INT-PUB 05-29}
\begin{document}
\title{Vanishing Bulk Viscosities and Conformal Invariance of the Unitary
Fermi Gas} 
\author{D.~T.~Son}
\affiliation{Institute for Nuclear
Theory, University of Washington, Seattle, Washington 98195-1550, USA}

\date{November 2005}
\begin{abstract}
  By requiring general-coordinate and conformal invariance of the
  hydrodynamic equations, we show that the unitary Fermi gas has zero
  bulk viscosity, $\zeta=0$, in the normal phase.  In the superfluid
  phase, two of the bulk viscosities have to vanish,
  $\zeta_1=\zeta_2=0$, while the third one $\zeta_3$ is allowed to be
  nonzero.
\end{abstract}
\maketitle



The unitary Fermi gas is a system where fermions of two components
interact through a zero-range two-body potential fine-tuned to
infinite scattering length.  This system is under active study, both
experimentally~\cite{Grimm,Jin,Ketterle,Salomon} and theoretically.
At zero temperature, the system is located midway between the 
Bose-Einstein condensation (BEC) and
the BCS regimes~\cite{Eagles,Leggett,Nozieres}.  The most remarkable
property of this system is the absence of any intrinsic dimensionful
parameter except the density and the temperature.  This implies, on
the one hand, a lack of any small perturbative parameter.  On the
other hand, the properties of the system are universal, i.e.,
independent of the microscopic details, which leads to great
simplification in many problems.

Several recent experiments have concentrated on dynamic properties of
the unitary Fermi gas.  The trap breathing modes was studied in
Ref.~\cite{Thomas,Grimm-collective,Thomas-breakdown,Thomas-2transitions},
in which it was observed that the damping rate reaches a minimum near
unitary.  The expansion of the gas as it is released from a trap was
studied in Refs.~\cite{OHara,Salomon-intE}, where a hydrodynamic
behavior was observed.  Such dynamic processes should depend, to one
or another degree, on the kinetic coefficients.  According to fluid
mechanics~\cite{LL6}, a normal gas has three kinetic coefficients: the
shear viscosity $\eta$, the bulk viscosity $\zeta$, and the thermal
conductivity $\kappa$.  This is also the situation for the Fermi gas
above the critical temperature.  Below the critical temperature, the
Fermi gas is in the superfluid phase, and the number of kinetic
coefficients is five~\cite{LL6,Khalatnikov}: $\eta$, $\kappa$, and
three bulk viscosities, normally denoted as $\zeta_1$, $\zeta_2$, and
$\zeta_3$.

The purpose of this Letter is to show that at unitarity (i.e., when the
scattering length is infinite), certain kinetic coefficients vanish
exactly. Namely, in the normal phase the bulk viscosity vanishes, and
in the superfluid phase two of three bulk viscosities, $\zeta_1$ and
$\zeta_2$, vanish.

To emphasize the nontriviality of this result, we note that it is by
no mean related to the well-known fact that the Boltzmann equation for
a classical monoatomic gas, with two-body collision, implies zero bulk
viscosity.  When three-body collisions are included, the Boltzmann
equation generically yields a nonzero bulk viscosity~\cite{LL10}.  In
our case, the bulk viscosities vanish even when the Boltzmann equation
cannot be used, but the result is not expected to hold outside the
unitarity regime.

In the unitarity regime, scale invariance has been used to derive
nontrivial relationships between thermodynamic observables~\cite{Ho}.
But scale invariance, \emph{per se}, does not imply the vanishing of
any kinetic coefficient.  It only implies that all kinetic
coefficients are homogeneous functions of the temperature $T$ and the
chemical potential $\mu$.  From scale invariance one can immediately
write down the scaling behavior of the shear viscosity $\eta$, the
bulk viscosity $\zeta$, and the thermal conductivity $\kappa$,
\begin{equation}
  \eta = \hbar n \tilde\eta\left(\frac T\mu\right) , ~
  \zeta = \hbar n \tilde\zeta\left(\frac T\mu\right) , ~
  \kappa = \frac{\hbar n}m
     \tilde\kappa\left(\frac T\mu\right),
\end{equation}
where $\tilde\eta$, $\tilde\zeta$, and $\tilde\kappa$ are
dimensionless function of the ratio of the temperature and the
chemical potential.  In the superfluid phase, instead of one function
$\tilde\zeta$ one has three functions $\tilde\zeta_1$,
$\tilde\zeta_2$, $\tilde\zeta_3$.

The result advertised above is therefore stronger than what simple
scaling arguments would imply.  Heuristically, the vanishing of the
bulk viscosities can be understood as follow.  Consider a blob of a
unitary Fermi gas that undergoes a uniform expansion, where the
velocity $\v$ at point $\x$ being $\v(\x)=c\x$, where $c$ is some
constant.  Because the unitary Fermi gas does not have any intrinsic
scale, it should remains in thermal equilibrium throughout the whole
process of uniform expansion.  This means entropy is not produced
during a flow with $\v=c\x$.  Looking at the equation for entropy
production, one finds $\zeta=0$.  Applying the same argument for a
blob of a superfluid unitary Fermi gas undergoing uniform expansion
with the same normal and superfluid velocity profile, $\v_n=\v_s=c\x$,
we find $\zeta_2=0$.  From the well-known inequality
$\zeta_1^2\leq\zeta_2\zeta_3$
one concludes $\zeta_1$ too has to
vanish, and only $\zeta_3$ can be nonzero.

Putting this argument on a more precise footing is, however, not
straightforward.  There is no regular solution to the hydrodynamic
equations that describe uniform expansion.  In this rest of this Letter
we put the heuristic argument above on a firm mathematical ground.  We
will show how the vanishing of $\zeta$ in the normal phase and
$\zeta_1$ and $\zeta_2$ follows from the requirement that the
hydrodynamic equation exhibits the conformal invariance of the
microscopic theory.

A general discussion of symmetries of the unitary Fermi gas is given
in Ref.~\cite{Son:2005rv}; for convenience we briefly review these
symmetries here.  Let us start by discussing the microscopic theory.
Due to the universality of the unitary Fermi gas, any short-range
two-body interaction can be used, if it corresponds to infinite
scattering length.  In particular, we can choose to work with
the following local Lagrangian (here and below $\hbar=1$),
\begin{equation}
  \L = i\psi^\+\d_t\psi - \frac1{2m}|\nabla\psi|^2 
       + q_0\psi^\+\psi\sigma
       - \frac12(\nabla\sigma)^2 - \frac{\sigma^2}{2r_0^2}\,, 
\end{equation}
which describe a system of fermions interacting through the Yukawa
potential $V(r)=-q_0^2 (4\pi r)^{-1} e^{-r/r_0}$.  Infinite scattering
length is achieved by requiring $mq_0^2r_0$ to be equal to a critical
number (whose numerical value is 21.1...).  Moreover $r_0$ and $q_0$
can be time-dependent; the only requirement is that $q_0^2r_0$ is held
fixed.  For universality we also need to keep $r_0(t)$ small compared
to any length scale in the problem.  In particular, if one perform the
following transformation
\begin{equation}\label{q0r0}
\begin{split}
  q_0(t) &\to q'_0(t) = \gamma(t) q_0(t), \\
  r_0(t) &\to r'_0(t) = \gamma^{-2}(t) r_0(t),
\end{split}
\end{equation}
then, from the point of view of the physics at length scale larger
than $r_0$, we have mapped our theory to itself, but not to another
theory, provided that $mq_0^2r_0$ is tuned to the value corresponding
to infinite scattering length.

The best way to expose the symmetries of the system is to put it in a
background gauge field $A_\mu$ and in a curved space with a 3D metric
tensor $g_{ij}(t,\x)$~\cite{Son:2005rv}.  The curved space here is a
not physical; it is simply a trick to get physical results.  The
action is now
\begin{multline}\label{S}
  S=\! \int\!dt\,d\x\,\sqrt g\biggl[ \frac i2 \psi^\+ \dlr_t \psi
    - \frac{g^{ij}}{2m}(\d_i+iA_i)\psi^\+(\d_j-iA_j)\psi \\
    + (q_0\sigma-A_0)\psi^\+\psi - \frac{g^{ij}}2\d_i\sigma\d_j\sigma
    - \frac{\sigma^2}{2r_0^2} \biggr],
\end{multline}
where $g=\det|g_{ij}|$ and $g^{ij}$ is the inverse metric of $g_{ij}$.
The action is invariant under the following infinitesimal
transformations: (i) gauge transformations, parameterized by a
function of space and time $\alpha(t,\x)$,
\begin{equation}\label{gauge}
  \delta\psi = i\alpha\psi, \qquad \delta A_0 = -\d_t \alpha,\qquad
  \delta A_i = -\d_i \alpha;
\end{equation}
(ii) local diffeomorphism,
parameterized by three functions of space and time $\xi^i(t,\x)$,
$i=1,2,3$, 
\begin{subequations}\label{gc}
\begin{align}
  \delta\psi &= -\xi^k\d_k\psi, \qquad
  \delta\sigma = -\xi^k\d_k\sigma, \\
  \delta A_0 &= -\xi^k\d_k A_0 - A_k \dot\xi^k ,\\
  \delta A_i &= -\xi^k\d_k A_i - A_k\d_i\xi^k + mg_{ik}\dot\xi^k,
  \label{nonrel-gci-Ai}\\
  \delta g_{ij} &= -\xi^k\d_k g_{ij} -g_{ik}\d_j\xi^k -
  g_{kj}\d_i\xi^k .
\end{align}
\end{subequations}
and (iii) ``conformal transformations,'' parameterized by a
function $\beta(t)$ of time only, 
\begin{equation}\label{conformal}
  \delta O = -\beta \dot O - \Delta[O] \dot\beta O,
\end{equation}
where $\Delta[O]$ is the dimension of the field $O$, defined so that
$\Delta[\psi]=\Delta[\sigma]=\frac34$, $\Delta[A_0]=1$,
$\Delta[A_i]=0$, and $\Delta[g_{ij}]=-1$.  
One can think about~(\ref{conformal}) as time reparameterization 
$t\to t'=t+\beta$.
By direct substitution one
can check that the action~(\ref{S}) is invariant under the
transformations~(\ref{gauge}) and (\ref{gc}), and
under~(\ref{conformal}) combined with~(\ref{q0r0}) with
$\gamma=1-\frac14\dot\beta$.  Galilean invariance is a combination of
(\ref{gauge}) and (\ref{gc}) with $\alpha=m{\bf V}\cdot\x$ and
$\xi^k=V^kt$, and scale invariance is a combination of (\ref{gc}) and
(\ref{conformal}) with $\beta=bt$ and $\xi^k=\frac12 bx^k$, where
$V^k$ and $b$ are constants.

The local diffeomorphism~(\ref{gc}) reduces to reparameterization of
space when $\xi^k$ is time independent: the transformation laws come
directly from the 3D tensor structure of the object under
consideration (i.e., scalar in the case of $\psi$, $\sigma$, and
$A_0$, vector in the case of $A_i$, and tensor in the case of
$g_{ij}$).  Equations~(\ref{gc}) extend this invariance to
\emph{time-dependent} diffeomorphisms.

In flat space $g_{ij}=\delta_{ij}$ and zero background fields
$A_0=A_i=0$, the long-distance, long-time dynamics of the system is
described by a set of hydrodynamic equations~\cite{LL6}.  This should
remain true when the external fields is turned on and when the space
is curved, if the following conditions are met: the fields are
sufficiently weak and vary over length and time scales much larger
than all microscopic scales; the curvature of space is small and
varies slowly, also compared to all microscopic scales.  The
hydrodynamic equation should then be properly modified to take into
account the external field and the curved metric.  In writing these
equations down, we require that the symmetries of the
microscopic theory are inherited by the hydrodynamic theory.  This
condition arises from the fact that the hydrodynamic equations, which
can be used to determine the response of the system on external
perturbations, imply concrete forms of the fully retarded Green's
functions of the hydrodynamic variables.  The Ward identities that
come from the symmetries~(\ref{gc}) should be satisfied by these fully
retarded Green's function, which is achieved if the hydrodynamic
equations have the same symmetries~\footnote{This argument fails for
the time-reversal symmetry, which maps fully retarded Green's
functions into fully advanced ones.}.

Our strategy is the following.  First we will modify the standard
hydrodynamic equations for the case of a nonzero external gauge field
and a curved metric.  There is a unique way to do so, as we shall see.
Then we will show that the resulting equations are inconsistent with
the conformal invariance unless certain kinetic coefficients vanish.

Consider the normal phase first.  We shall write the hydrodynamic
equations in term of the local mass density $\rho$, the local velocity
$v^i$, and the local entropy per unit mass $s$.  The set consists of
the continuity equation,
\begin{equation}\label{continuity}
  \frac1{\sqrt g}\d_t (\sqrt g\,\rho) + \nabla_i (\rho v^i) = 0,
\end{equation}
the equation of momentum conservation,
\begin{equation}\label{momentum-conservation}
  \frac1{\sqrt g}\d_t (\sqrt g\, \rho v_i) + \nabla_k \Pi^k_i 
    = \frac\rho m (E_i - F_{ik} v^k),
\end{equation}
and the entropy production equation
\begin{equation}\label{S-prod}
  \frac1{\sqrt g}\d_t(\sqrt g\,\rho s) 
  + \nabla_i\Bigl(\rho v^i\d_i s - \frac\kappa T\d^i T\Bigr) = \frac{2R}T\,.
\end{equation}
We follow closely the notations of Ref.~\cite{LL6}: $\Pi_{ik}$ is the
stress tensor, $\kappa$ is the thermal conductivity, and $R$ is the
dissipative function.  The obvious modifications are the replacement
of the derivatives $\d_i$ by the covariant derivatives
$\nabla_i$~\cite{LL2} and the appearance of the force term in the
momentum conservation equation~(\ref{momentum-conservation}), which
comes from the electric ($E_i=\d_t A_i-\d_i A_0$) force and the
magnetic $F_{ik}=\d_i A_k-\d_k A_i$) Lorentz force. The stress tensor
can be written as
\begin{equation}
  \Pi_{ik} = \rho v_i v_k + p g_{ik} - \sigma'_{ik},
\end{equation}
where $p$ is the pressure and $\sigma'_{ik}$ is the viscous stress
tensor.  The information about the kinetic coefficients are contained
in $\sigma'_{ik}$ and $R$.

Consider the dissipationless limit first, setting $\sigma'=R=0$.  One
can check that the hydrodynamic equations are invariant with respect
to the general-coordinate transformations, provided that $A_0$, $A_i$,
and $g_{ij}$ transform as in Eqs.~(\ref{gc}), and $\rho$, $s$, and
$v^i$ transform as
\begin{align}
  \delta\rho &= -\xi^k\d_k\rho, \qquad 
  \delta s = -\xi^k\d_k s, \\
  \delta v^i &= -\xi^k\d_k v^i + v^k\d_k\xi^i + \dot \xi^i.
\end{align}
The transformation law for $v^i$ contains terms coming from the vector
nature of $v^i$, but also a $\dot \xi^i$ term that can be understood
if one recall $\v$ changes under Galilean boosts ($\xi^i=V^it$):
$\v\to\v+{\bf V}$

Now consider the dissipative terms.  To keep the equation consistent
with diffeomorphism invariance, one must require that $\sigma'_{ik}$
and $R$ transform as a two-index tensor and a scalar, respectively,
\begin{align}
  \delta\sigma'_{ij} &= -\xi^k\d_k\sigma'_{ij} - \sigma_{kj}\d_i\xi^k
  -\sigma_{ik}\d_j\xi^k, \label{deltasigma} \\
  \delta R &= -\xi^k \d_k R \,.
\end{align}
In flat space the viscous stress tensor is given by
\begin{equation}
  \sigma'_{ij} = \eta(\d_i v_j + \d_j v_i) + \left(\zeta-\tfrac23\eta\right)
  \delta_{ij} \d_k v^k .
\end{equation}
The naive extension to curve space,
\begin{equation}
  \sigma'^{\rm naive}_{ij} = \eta(\nabla_i v_j + \nabla_j v_i)
  +\left(\zeta - \tfrac23 \eta\right) g_{ij} \nabla_k v^k,
\end{equation}
is, however, not a pure two-index tensor: its variation under
diffeomorphism contains extra terms proportional to $\dot\xi^k$:
\begin{equation}
\begin{split}
  \delta \sigma'^{\rm naive}_{ij} &= 
  \eta[\nabla_i(g_{jk}\dot\xi^k) + \nabla_j(g_{ik}\dot\xi^k) ]\\
  &\quad+\left(\zeta - \tfrac23 \eta\right) g_{ij} \nabla_k \dot\xi^k\\
  &\quad+ \textrm{terms in Eq.~(\ref{deltasigma}}).
\end{split}
\end{equation}
These terms can be canceled out by adding terms proportional to the
derivatives of the metric tensor to $\sigma'_{ik}$.  Limiting oneself
to terms containing the least number of derivatives, the terms needed
are determined uniquely,
\begin{equation}
  \sigma'_{ij} = \eta(\nabla_i v_j + \nabla_j v_i +\dot g_{ij})
  +\left(\zeta - \frac23 \eta\right) g_{ij} \left(\nabla_k v^k
  + \frac{\dot g}{2g} \right).
\end{equation}
Now one can check that $\sigma'_{ik}$ transforms according to
Eq.~(\ref{deltasigma}).

Similarly, the dissipative function $R$ becomes, in curved space
\begin{multline}
  2R =
  \frac\eta 2 \left(\nabla_i v_j + \nabla_j v_i 
  - \frac23 g_{ij}\nabla_k v^k + \dot g_{ij} 
  - \frac13 g_{ij}\frac{\dot g}g \right)^2\\
  + \zeta \left(\nabla_i v^i + \frac{\dot g}{2g}\right)^2
  + \frac\kappa T\d_i T\d^i T .
\end{multline}

Let us now specialize on Fermi systems at infinite scattering length,
and discuss its conformal invariance.  The dissipationless
hydrodynamic equation is invariant under~(\ref{conformal}), if the
dimensions of different fields are
\begin{equation}
  \Delta[\rho] = 2\Delta[\psi]= \tfrac32\,, 
  \quad \Delta[s]=0, \quad \Delta[v^i]=1.
\end{equation}
Now let us consider the dissipation terms.  From dimensional analysis
one find that one have to set
\begin{equation}
  \Delta[\eta] = \Delta[\zeta] = \Delta[\kappa] = \tfrac32\,,
\end{equation}
for the hydrodynamic equation to be scale-invariant.  However,
conformal invariance is not preserved generically.  The culprit is the
$\dot g_{ij}$ that transform as
\begin{equation}
  \delta \dot g_{ij} = -\beta\ddot g_{ij} + \ddot\beta g_{ij}\,,
\end{equation}
which leads to $\sigma'_{ij}$ and $R$ not to conform to the pattern
of~(\ref{conformal}),
\begin{subequations}\label{deltasigmaR}
\begin{align}
  \delta\sigma'_{ij} &= -\beta \dot\sigma'_{ij} -\tfrac32\dot\beta 
  \sigma'_{ij} + \tfrac32\zeta \ddot\beta g_{ij}\,,\\
  \delta R &= -\beta\dot R - \frac72 \dot\beta R
  +\frac32\zeta\ddot\beta \left( \nabla_i v^i + \frac{\dot g}{2g}\right),
\end{align}
\end{subequations}
unless the bulk viscosity $\zeta$ vanishes.  Thus the requirement of
conformal invariance of the hydrodynamic equations implies
$\zeta=0$. [Had we known only the scale invariance, i.e., had we
restricted $\beta(t)$ to be of the form $\beta(t)=bt$, the
$\ddot\beta$ terms would have been absent from
Eqs.~(\ref{deltasigmaR}) and we would have been unable to reach this
conclusion.]

Similarly, we can repeat the argument for the superfluid case.  The
hydrodynamics of superfluids contains an additional degree of freedom,
which is the condensate phase $\varphi$, whose gauge-covariant
gradient is the superfluid velocity,
\begin{equation}\label{vs}
  v^s_i = \frac\hbar m (\d_i\varphi + A_i)\,.
\end{equation}
It transforms in the same way as the normal velocity $v_i \equiv
v^n_i$ under general-coordinate and conformal transformations.  A
consequence is that the relative velocity between the superfluid and
the normal component $w^i=v_s^i-v^i$ transforms as a pure vector under
diffeomorphism,
\begin{equation}
  \delta w^i = -\xi^k\d_k w^i + w^k\d_k \xi^i.
\end{equation}
The $\dot\xi^i$ term in the variation cancels between $\delta v_s$
and $\delta v$.

The diffeomorphism-invariant dissipative function in curved space is
\begin{multline}
 2R =  \frac\eta 2 \left(\nabla_i v_j + \nabla_j v_i 
  - \frac23 g_{ij}\nabla_k v^k + \dot g_{ij} 
  - \frac13 g_{ij}\frac{\dot g}g \right)^2\\
  + 2\zeta_1  \left(\nabla_i v^i + \frac{\dot g}{2g}\right)
    \nabla_j (\rho_s w^j)
  + \zeta_2 \left(\nabla_i v^i + \frac{\dot g}{2g}\right)^2\\
  + \zeta_3 [\nabla_i(\rho_s w^i)]^2
  + \frac\kappa T \d_i T\d^iT .
\end{multline}
Under conformal transformations, $R$ transforms as
\begin{equation}
  \delta R = -\beta\dot R - \frac72 \dot\beta R
  +\frac32\zeta_1\ddot\beta \nabla_i(\rho w^i)
  +\frac32\zeta_2\ddot\beta \left( \nabla_i v^i + \frac{\dot g}{2g}\right).
\end{equation}
The requirement of conformal invariance of superfluid hydrodynamics
implies that the $\ddot\beta$ terms must have vanishing coefficients,
i.e., $\zeta_1=\zeta_2=0$.

In conclusion, we find that in the unitary limit the bulk viscosity
vanishes in the normal phase.  In the superfluid phase two of the
three bulk viscosities vanishes.  This vanishing of the bulk
viscosities are directly related to the conformal invariance of the
unitary Fermi gas.


It should be possible to check the result derived in this Letter by
using the Boltzmann equation in the two regimes where it applies: the
high-temperature regime $T\gg\mu$ and the low temperature regime
$T\ll\mu$.  In the intermediate regime $T\sim\mu$ the Boltzmann
equation is not reliable, since we do not have weakly coupled
quasiparticles.  The result derived in this paper, however, should be
valid for all regimes of $T/\mu$.

With respect to the recent experimental
findings~\cite{Thomas,Grimm-collective,Thomas-breakdown,Thomas-2transitions}
it is tempting to speculate that the dip in the damping rate of the
radial breathing modes near the Feshbach resonance could
\emph{partially} be due to the vanishing of the bulk viscosities at
unitarity.  Another source for the reduction of damping is probably
the decrease of the shear viscosity and the thermal conductivity at
strong coupling.  A study of the breathing modes using two-fluid
dissipative hydrodynamics may enable one to extract kinetic
coefficients from experimental data, and ultimately verify the results
derived in this Letter.

This work is supported, in part, by DOE Grant No.\
DE-FG02-00ER41132.


\begin{thebibliography}{99}

\bibitem{Grimm}
M.~Bartenstein \emph{et. al.,}
  Phys.\ Rev.\ Lett.\ {\bf 92}, 120401 (2004).

\bibitem{Jin}
  C.~A.~Regal, M.~Greiner, and D.~S.~Jin,
  Phys.\ Rev.\ Lett.\ {\bf 92}, 040403 (2004).

\bibitem{Ketterle}
  M.~W.~Zwierlein \emph{et.al.,}  
  Phys.\ Rev.\ Lett.\ {\bf 92}, 120403 (2004).

\bibitem{Salomon}
  T.~Bourdel \emph{et. al.,}
  Phys.\ Rev.\ Lett.\ {\bf 93}, 050401 (2004).

\bibitem{Eagles}
  D.~M.~Eagles,
  Phys.\ Rev.\ {\bf 186}, 456 (1969).

\bibitem{Leggett}
  A.~J.~Leggett, 
in \emph{Modern Trends in the Theory of Condensed
  Matter} (Springer, New York, 1980).

\bibitem{Nozieres}
  P.~Nozi\`eres and S.~Schmitt-Rink,
  J.\ Low Temp.\ Phys.\ {\bf 59}, 195 (1985).

\bibitem{Thomas}
 J.~Kinast, 
  S.~L.~Hemmer, M.~E.~Gehm, A.~Turlapov, and J.~E.~Thomas,
  Phys.\ Rev.\ Lett.\ {\bf 92}, 150402 (2004).

\bibitem{Grimm-collective}
M.~Bartenstein \emph{et. al.,}
Phys.\ Rev.\ Lett.\ {\bf 92}, 203201 (2004).

\bibitem{Thomas-breakdown}
J.~Kinast, A.~Turlapov, and J.~E.~Thomas,
Phys.\ Rev.\ A {\bf 70}, 051401(R) (2004); {\bf 71}, 029901(E) (2005).

\bibitem{Thomas-2transitions}
J.~Kinast, A.~Turlapov, and J.~E.~Thomas,
Phys.\ Rev.\ Lett. {\bf 94}, 170404 (20005).

\bibitem{OHara}
K.~M.~O'Hara \emph{et. al.,}
Science {\bf 298}, 2179 (2002).

\bibitem{Salomon-intE}
T.~Bourdel \emph{et. al.,}
Phys.\ Rev.\ Lett.\ {\bf 91}, 020402 (2003).

\bibitem{LL6}
L.~D.~Landau and E.~M.~Lifshitz, \emph{Fluid Mechanics}
(Pergamon, New York, 1987), 2nd ed.

\bibitem{Khalatnikov}
I.~M.~Khalatnikov, {\em An Introduction to the Theory of Superfluidity} 
(Benjamin, New York, 1965).

\bibitem{LL10}
E.~M.~Lifshitz and L.~P.~Pitaevskii,
\emph{Physical Kinetics}
(Pergamon Press, New York, 1981).

\bibitem{Ho} See, e.g., the discussion in T.-L.~Ho, 
Phys.\ Rev.\ Lett.\ {\bf 92}, 090402 (2004).

\bibitem{Son:2005rv}
D.~T.~Son and M.~Wingate,
Ann.\ Phys.\ (N.Y.) {\bf 321}, 197 (2006).

\bibitem{LL2}
L.~D.~Landau and E.~M.~Lifshitz,
\emph{The Classical Theory of Fields}
(Pergamon, New York, 1975), 4th ed.

\end{thebibliography}
\end{document}